\documentclass{llncs}

\newif\ifdraft
\drafttrue

\usepackage[T1]{fontenc}
\usepackage[utf8]{inputenc}
\usepackage{textcomp}

\ifdraft
\usepackage[show]{ed}
\else
\usepackage[hide]{ed}
\usepackage{microtype}
\fi

\usepackage[firstinits=true,mincrossrefs=2,minnames=3,maxnames=3]{biblatex}
\bibliography{kwarc}
\AtBeginBibliography{%
\setcounter{maxnames}{100}
}

\usepackage{paralist}
\usepackage{csquotes}
\usepackage{courier}
\usepackage{amsmath}
\usepackage{amstext}
\usepackage{fancyvrb}
\usepackage{listings}
\usepackage{relsize}

\usepackage{tikz}

\usepackage{wasysym}

\usepackage{semantic-markup}
\usepackage{local}

\def\thetitle{Towards OpenMath Content Dictionaries as Linked Data}

\definecolor{NavyBlue}{cmyk}{0.94,0.54,0,0.3}
\usepackage[pdftex,pdfstartview=FitV,plainpages=false,pdfpagelabels,colorlinks=true,linkcolor=NavyBlue,citecolor=NavyBlue,urlcolor=NavyBlue,hypertexnames=true]{hyperref}
\hypersetup{%
  pdfauthor = {Christoph Lange},%
  pdftitle = {\thetitle},%
  pdfkeywords = {}%
}
\usepackage{url}

\usetikzlibrary{shapes}

\title{\thetitle} \author{Christoph Lange} \institute{Computer Science, Jacobs University
  Bremen,\\ \email{ch.lange@jacobs-university.de}}

\begin{document}

\maketitle

\begin{abstract}
  \enquote{The term \enquote{Linked Data} refers to a set of best practices for publishing and connecting structured data on the web}~\cite{BHB:LinkedDataStory09}.  Linked Data make the Semantic Web work practically, which means that information can be retrieved without complicated lookup mechanisms, that a lightweight semantics enables scalable reasoning, and that the decentral nature of the Web is respected.  OpenMath Content Dictionaries (CDs) have the same characteristics -- \emph{in principle}, but not yet in practice.

The Linking Open Data movement has made a considerable practical impact: Governments, broadcasting stations, scientific publishers, and many more actors are already contributing to the \enquote{Web of Data}.  Queries can be answered in a distributed way, and services aggregating data from different sources are replacing hard-coded mashups.  However, these services are currently entirely lacking mathematical functionality.  I will discuss real-world scenarios, where today's RDF-based Linked Data do not quite get their job done, but where an integration of OpenMath \emph{would} help -- were it not for certain conceptual and practical restrictions.

I will point out conceptual shortcomings in the OpenMath 2 specification and common bad practices in publishing CDs and then propose concrete steps to overcome them and to  contribute OpenMath CDs to the Web of Data.
\end{abstract}

\section{Linked Data State of the Art}
\label{sec:state}

The Linked Data principles, established by \person{Berners-Lee} in 2006~\cite{TBL:LinkedData06} consist of four simple rules for publishing machine-understandable data on the web\footnote{here cited as paraphrased by Wikipedia~\cite{wikipedia:linked-data}}:

\begin{enumerate}
\item Use URIs to identify things.
\item Use HTTP URIs so that these things can be referred to and looked up (\enquote{dereferenced}) by people and user agents.\footnote{I.\,e., the URI is treated as a URL.}
\item Provide useful\footnote{This usually means: machine-understandable.} information about the thing when its URI is dereferenced, using standard formats such as RDF/XML.
\item Include links to other, related URIs in the exposed data to improve discovery of other related information on the Web.
\end{enumerate}

\begin{figure}[htb]
  \pgfimage[width=\textwidth]{\KWARCpic{semweb}{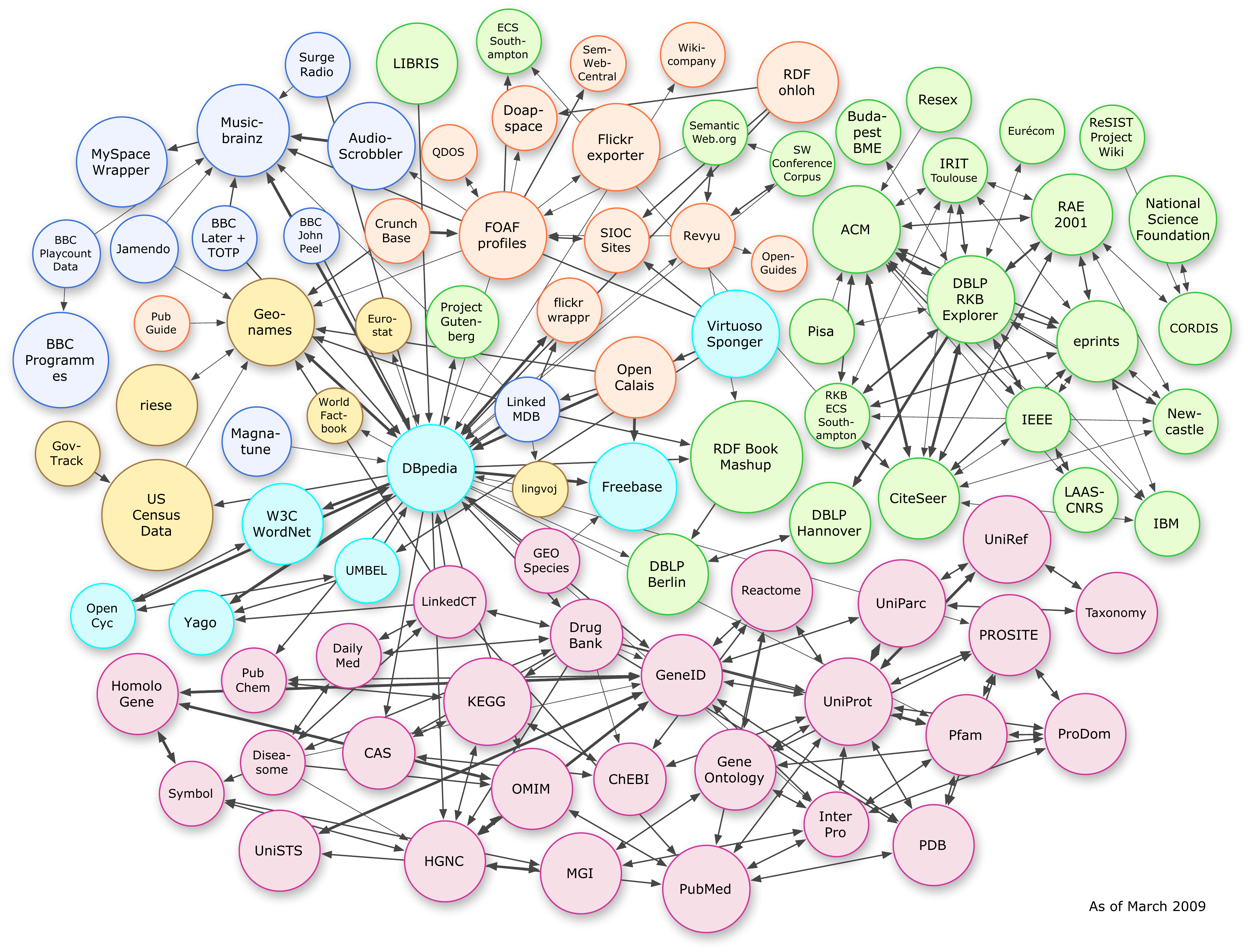}}
  \caption{Linked Open Datasets as of March 2009~\cite{LOD-cloud}}
  \label{fig:lod}
\end{figure}

These principles are widely considered to have made the Semantic Web vision work practically.  A lot of providers have already published their data according to these principles and interlinked them with other datasets (cf.\ figure~\ref{fig:lod}).  The hub in this big picture is DBpedia, a huge collection of general-purpose data extracted from Wikipedia and made available as RDF.  Data from specific domains, such as scientific publications (green), biomedicine (pink), social networks (orange), multimedia (dark blue), and government statistics (yellow) have also been published as Linked Open Data.  Linked Data do not \emph{have} to be open\footnote{In fact they can also be useful in intranet settings, cf.~\cite{Servant:LinkingEnterpriseData08}}, but making datasets open of course helps to interlink and reuse knowledge; therefore, the open datasets have so far been the most visible and most widely used instance of Linked Data.  Applications include browsers, which allow users to traverse the Web of Data and discover connections, semantic search engines and indexes, which enable a more accurate information retrieval than keyword-based engines, as well as mashups that aggregate Linked Data from distributed sources and expose them via a coherent user interface (see, e.\,g., \cite{HMF:ResearchersMap09} for an interactive map of database researchers and their publications, filterable by research topics).

\section{The Need for Mathematical Semantics}
\label{sec:applications}

None of the Linked Open datasets and applications known to date deals with mathematical knowledge, not counting mere descriptions that do not involve any mathematical semantics, e.\,g., of mathematical publications or mathematical research topics, as they can be found in publication datasets or DBpedia.  With \enquote{Linked Open Numbers}~\cite{VKRL:LinkedOpenNumbers10}, one mathematical dataset has been published, but that was not to be taken serious.  There, every natural number from 1 to 999,999,999 is described with its predecessor, successor, natural logarithm, and its name in various natural languages.  This is pretty useless information\footnote{On the other hand, it \emph{might} be useful to publish as Linked Data facts about numbers that are \emph{hard} to compute, e.\,g.\ factorizations of large numbers.}, and indeed \enquote{Linked Open Numbers} was an April fool's joke cartooning the rampant bad habit of mindlessly publishing datasets that are very large but not reasonable at all.

\begin{lstlisting}[language=Turtle,float,label=lst:geese,caption={Geese on the Isle of Wight, RDF data in Turtle notation, from \url{data.gov.uk} (all URIs abbreviated, namespace prefix mappings omitted for brevity; see~\cite{VLHBD:SemGovStatData10} for full example)}]
ahs:EH100                            # just some ID for this data point
  scv:dimension env:isle-of-wight ;  # the "region" dimension
  scv:dimension env:year-2008 ;      # the "time" dimension
  scv:dimension env:geese ;          # the type of items counted
  rdf:value     "693"^^xsd:decimal ; # the count
  scv:dataset   ahs2:livestock .     # back-reference to the dataset
\end{lstlisting}

There is, however, no doubt, that mathematical semantics is needed in order to improve, or even enable, certain \emph{serious} applications of Linked Data.  I consider statistical datasets, which are now being published as RDF Linked Data, e.\,g., by the UK and US governments, a prime example.  \person{Omitola} et al.\ have, for example, used such data in order to answer queries for public sector information in the user's home region by aggregating data about, e.\,g., political representatives of the local constituencies, crime statistics for the local county, and waiting list statistics of local hospitals~\cite{OmitolaEtAl:PostCodeDataCaseStudy10}.  At the moment, these datasets contain a lot of data points (e.\,g.\ the number of geese on the Isle of Wight in 2008; cf.\ listing~\ref{lst:geese}), without making their origin semantically explicit.  We have proposed an extension of the relevant Statistical Core Vocabulary (SCOVO), which allows to express the latter knowledge, saying, e.\,g., \enquote{the things that we are counting here are geese (e.\,g.\ by referencing \url{http://dbpedia.org/resource/Goose}) per area and per year}~\cite{VLHBD:SemGovStatData10}.  Mathematical knowledge becomes relevant when modeling \emph{derived values}, such as the geese population density of a region in a given year, defined as the number of geese divided by area.\footnote{The geese population density is a fictitious example, but in the actual datasets, there are derived values such as the [human] population density of various census regions, or the average number of jobs per citizen.}  At the moment, there are a lot of derived values in the datasets published, simply given as additional raw data points.  For a client consuming these data, there is no way of verifying their correctness or applying the same derivation rule to new or changed base values, because the derivation rule is not made explicit.  We have shown how to make their mathematical semantics explicit – first on the instance level, as that integrates most easily into existing datasets.  Let the data point with the ID \textit{ahs:AR100} be the area of the Isle of Wight, and let \textit{ahs:PD100} be the geese population density of the Isle of Wight in 2008, then we could express the fact that the latter is \textit{ahs:EH100} divided by \textit{ahs:PD100} by referencing the OpenMath symbol for division (cf.\ listing~\ref{lst:density}).  In a second step, the same could be done on vocabulary level: In addition to, or alternatively to, explicitly representing the derivation of each data point, one could model a general rule that \enquote{for each data point $p$ containing a \enquote{population} of some region $r$ at some point $t$ in time and for each data point $a$ containing the area of $r$ [at time $t$], the population density $d$ of $r$ at time $t$ is defined as $d:=\frac{p}{a}$}.  Recall, however, that the semantics of Linked Data vocabularies is usually intentionally weak in order to enable large-scale applications.  Such general rules would require more powerful clients and query engines and might therefore not work as universally as semantically more lightweight (albeit blatantly redundant) annotations of individual data points.

\begin{lstlisting}[language=Turtle,float,label=lst:density,caption={Geese population density of the Isle of Wight, with its mathematical semantics}]
# the density is computed by ...
ahs:PD100 sl:computedFrom [
  # ... calling OpenMath's arith1#divide
  sl:function <http://www.openmath.org/cd/arith1#divide> ;
  sl:arguments
    # ... passing the value of the EH100 data point as first argument
    [ sl:argPosition "1"^^xsd:int ;
      sl:argValue ahs:EH100 ] , 
    # ... and the value of the AR100 data point as second argument
    [ sl:argPosition "2"^^xsd:int ;
      sl:argValue ahs:AR100 ] ].
\end{lstlisting}

For computing such a derivation, a Linked Data client has to translate these RDF data to an OpenMath object, which has to be fed to a computation service, e.\,g.\ a service that speaks SCSCP~\cite{FHKLR:SCSCP08,SCSCP}.  We have detailed the translation in~\cite{VLHBD:SemGovStatData10}.  For standard symbols, such as \textit{arith1\#divide} here, the translation is pretty straightforward.  Computing the division should not be a problem for any OpenMath-aware service, as there is certainly a phrasebook mapping \textit{arith1\#divide} to the native division operator of some computer algebra system.

But now suppose that there are more complex, non-standard derivations in our statistical dataset.  This makes the case for publishing OpenMath CDs as Linked Data, by the following considerations: Suppose the dataset contains the Human Development Index (HDI) of a country\footnote{\url{http://en.wikipedia.org/wiki/Human_Development_Index}}.  Assuming that the four required auxiliary data points have already been computed (\textit{LE} = life expectancy index, \textit{ALI} = adult literacy index, \textit{GEI} = gross enrollment index, and \textit{GDP} = an index computed from the gross domestic product per capita at purchasing power parity, all normalized to a scale between $0$ and $1$), the \textit{HDI} is defined as $\frac{1}{3}(\mathit{LE}+\frac{2}{3}\mathit{ALI}+\frac{1}{3}\mathit{GEI}+\mathit{GDP})$.  In~\cite{VLHBD:SemGovStatData10}, we propose that the dataset publishers define the \textit{HDI} and its derivation as a symbol in an OpenMath CD that accompanies the dataset, e.\,g.\  \url{http://example.org/statistics}.  Now suppose there is a derived data point annotated as \lstinline[language=Turtle]|sl:computedFrom [ sl:function <http://example.org/|\\
\lstinline[language=Turtle]|statistics#hdi> ; ... ]| in analogy to listing~\ref{lst:density}.  As OpenMath-based computation services and thus phrasebooks are developed independently from datasets being published, we have little to no chance to expect a phrasebook supporting the \url{http://example.org/statistics} CD.  Therefore, we propose to add support for processing OpenMath CDs to Linked Data clients.  For (re)computing an HDI data point derived from four other data points containing the \textit{LE}, \textit{ALI}, \textit{GEI}, and \textit{GDP} values, the client would download the definition of the \url{http://example.org/statistics#hdi} symbol from the CD, expand the mathematical expression using the definition, and then send that expanded expression, which only uses operators from the universally understood \textit{arith1} CD, to the computation service.\footnote{Here, we assume that those values, from which the HDI is computed, are either hard-coded in the dataset, or that they have been computed before, using the same method.}

So far, I have outlined one use case, where OpenMath CDs as Linked Data would be needed.  In the following section, I will point out what actions on the OpenMath side that requires.  Note that, while the Linked Data principles have been devised in the context of RDF, and while all contemporary Linked Open datasets are available as RDF, the Linked Data guidelines do not prescribe RDF.  In fact, RDF might not be the most appropriate representation for mathematical objects.  It is at least quite cumbersome to break the ordered tree structure of mathematical expressions down to unordered RDF triples (cf.\ \cite{Marchiori:MathematicalSemanticWeb03} for one never-adopted suggestion on how that could be done, and \cite{VLHBD:SemGovStatData10} for a critical review).  For the remainder of this paper, I assume that CDs will be published in their reference XML encoding.

\section{Linked Data Principles in OpenMath}
\label{sec:openmath}

First, let us see how much the Linked Data principles cited in section~\ref{sec:state} are already respected in the practice of publishing OpenMath CDs:

\begin{enumerate}
\item\label{it:om-uri} Hardly any CD author uses \textit{CDBase}, which indicates a lack of awareness that things \emph{can} be identified by URIs.
\item The URIs used for OpenMath CDs/symbols are always HTTP URLs, but due to the inconsequent usage of \textit{CDBase} (cf.\
 principle \ref{it:om-uri}), most published CDs have the base URI \url{http://www.openmath.org/cd} (i.\,e.\ the default value for \textit{CDBase}).  Even when disregarding the following principles, this is at least bad style, as the authors who publish such CDs usually do not have control over the \url{openmath.org} domain.
\item\label{it:om-dereference} If people are aware of the fact that OpenMath CDs and symbols have a URI, they usually merely consider it a globally unique \emph{name}, but not a means of locating information about a CD or symbol.  At \url{openmath.org}, the URIs/URLs of CDs at least redirect to human-readable HTML renderings (e.\,g.\ \url{http://www.openmath.org/cd/arith1#plus}$\leadsto$\url{http://www.openmath.org/cd/arith1.xhtml#plus}), but that does not help a \emph{machine} that is interested in a description of a symbol:  Neither is the HTML semantically annotated (e.\,g.\ with RDFa, or with parallel markup in the case of rendered MathML formulæ), nor is the CD in its original XML encoding available from that URI in a way that would not require human brain-power.
\item OpenMath CDs are not integrated into the Web of Data at all.  Some of the standard CDs make references to mathematical literature, e.\,g.\ sections of the Handbook of Mathematical Functions by \person{Abramowitz} and \person{Stegun}~\cite{AbrSte:hmf64}.  Hyperlinks to its digital counterpart, the Digital Library of Mathematical Functions (DLMF~\cite{DLMF}), would be more appropriate here.\footnote{By the same argument as for principle~\ref{it:om-dereference}, the DLMF content is merely machine-\emph{readable}, but not machine-\emph{understandable}, but, on the other hand, the \person{Abramowitz}/\person{Stegun} book is even only \emph{human-readable}.}  Other than that, I am not aware of other links in CDs.  However, links to background information about certain mathematical operators or functions, e.\,g.\ to the DBpedia editions of their Wikipedia articles, would make sense.  Conversely, backlinks from DBpedia to OpenMath CDs would make sense (\enquote{go there if you want a description of the mathematical semantics of \url{http://dbpedia.org/resource/Logarithm}}).
\end{enumerate}

This lack of compliance with the Linked Data principles is not completely to blame on bad habits among the publishers of OpenMath CDs; it is also caused by technical and even conceptual problems:

\begin{itemize}
\item No MIME type for OpenMath objects/CDs has been specified.  When publishing Linked Data, it is good practice to make both machine- and human-readable descriptions of the same things available from their URIs (cf.\ principle~\ref{it:om-dereference} and~\cite{BCH:pldow07}).  That is, from the URI of an OpenMath CD, \emph{both} the CD XML file \emph{and} its human-friendly HTML rendering should be available.  We could even make an RDF description of the CD available, as RDF is most widely understood by Linked Data clients, and as we have machinery for translating OpenMath CDs to RDF (cf.~\cite{Lange:omwiki09}).  The client indicates the desired data encoding by requesting a particular MIME type in the \textit{Accept} header of its  HTTP request; this mechanism is called \emph{content negotiation} (cf.~\cite{W3C:CoolURIs}).  In accordance with best practices, I suggest \textit{application/openmath+xml} to be introduced for the XML encoding of OpenMath.\footnote{It is subject to further discussion in the community whether the same MIME type should be used both for OpenMath objects and CDs.}
\item CDs are not really meant as \emph{machine}-understandable descriptions of symbols.  They are mainly intended as somewhat rigorous descriptions for those \emph{humans} who implement phrasebooks, i.\,e.\ translations between OpenMath Objects and the native languages of, e.\,g., computer algebra systems.  This view is encouraged by the OpenMath 2.0 specification, which says \enquote{It is important to stress that it is not Content Dictionaries themselves which are being transmitted, but some \enquote{mathematics} whose definitions are held within the Content Dictionaries.}~\cite{BusCapCar:2oms04}  This view is plainly wrong on the Web of Data!  By the \enquote{meta} CD, there is at least a well-meant approach to communicating CDs as OpenMath Objects\footnote{… which even comes into operation in SCSCP~\cite{SCSCP}}.  Most OpenMath-aware software, except a few editors, usually only supports OpenMath objects (= formulæ), but not CDs.  For Linked Data applications, that has to change, and actually that change is not hard to make, because the CD XML format is well-specified and easy to implement.
\item The semantics of FMPs is too weak.  The application scenario outlined in section~\ref{sec:applications} assumes that FMPs carry definitions of symbols, but FMPs are not required to do so, as they could also carry asserted properties of symbols.  Definitional FMPs have been discussed throughout the last 10 years (see, e.\,g., \cite{CO:CommProofIntMathDoc00}) but still have not made it into the OpenMath standard.  We might even need to go one step further\footnote{according to personal communication with \person{Michael Kohlhase}} and introduce a notion of \emph{computational} FMPs, as, for example, implicit definitions are not useful for term rewriting either.  On the other hand, studying the practice of RDF-based Linked Data gives some hope, as, there, certain vocabulary terms are also used more liberally than they have been specified.  The \textit{rdfs:seeAlso} relation is semantically very weak (\enquote{used to indicate a resource that might provide additional information about the subject resource}~\cite{BrGu04:rdfs}), but when used with Linked Data, it is commonly assumed that it points at a URI that is again machine-comprehensible and contains further Linked-Data-compliant information.  Conversely, the \textit{owl:sameAs} relation is commonly used to declare that two things, despite having different URIs, are the same (cf.~\cite{W3C04:owl-guide}) – but hardly any Linked Data application makes use of the rest of the description logic based OWL ontology language, where this relation comes from.  Similarly, there are practical applications of OpenMath CDs, such as \person{Stratford}'s and \person{Davenport}'s unit converter~\cite{SD:UnitKnowledgeMgmt08}, which simply \emph{assume} that FMPs having the current symbol as the first argument of \textit{relation1\#eq} are definitional.
\item There is no mechanism for linking OpenMath symbols to anything else but other OpenMath symbols (e.\,g.\ to DBpedia data).  The latter is done via FMPs, but for the former we would have to be able to create typed links from OpenMath symbols to arbitrary URIs.  If the format restrictions for OpenMath symbol URIs were overcome (see next item), this \emph{could} be done by encoding RDF links as FMPs, as we have proposed in~\cite{LK:MathOntoAuthDoc09}.  Alternatively, one could permit RDFa metadata (a syntax for directly embedding RDF into XML-based languages) in CDs (see also~\cite{LK:MathOntoAuthDoc09}), but that would be a more intrusive change of the CD format.
\item From a Linked Data point of view, the OpenMath schema of symbol URIs being constructed as \texttt{cdbase / cd \# name} is too restrictive and should be liberalized.  Not only are there \enquote{legacy} URIs having other formats out there on the Web of Data, but also every data publisher may have good reasons not to choose \enquote{hash URIs} (see, e.\,g., \cite{W3C:VocabPub}).  As processing everything after the \texttt{\#} is up to the client, the consequent use of hash URIs for OpenMath symbols forces clients to always download a complete CD from the server, in which they would then have to locate the symbol with the desired name (e.\,g.\ using the \texttt{/CD/CDDefinition[Name = ...]} XPath expression).  However, for a large CD, of which a client is only interested in few symbols, it would be more efficient to use \enquote{slash URIs}, such as \url{http://cdba.se/cd/name}, or similar formats, such as the MMT URIs under development for OMDoc~\cite{KohRabZho:tmlmrsca10}.  In my opinion, it was also wrong to impose OpenMath's strict \texttt{cdbase / cd \# name} schema on Content MathML by de facto deprecating the liberal \texttt{csymbol/@definitionURL} attribute, which is now only permitted in non-strict markup~\cite{CarlisleEd:MathML09}.
\end{itemize}

I suggest that the OpenMath 3 specification address the conceptual issues and provide practical guidelines on how to address the technical issues.

\section{Conclusion and Future Work}
\label{sec:conc}

Looking at the state of the art of Linked Data and Linked Open datasets, we have identified a lack of mathematical semantics.  We have pointed out how applications would benefit from mathematically annotated Linked Data and suggested OpenMath CDs, in combination with the prevalent RDF, to be used for that purpose.  That, however, poses a number of technical and conceptual requirements on the OpenMath community, which I have described in detail, and which should be addressed in OpenMath 3.

\begin{figure}[htb]
  \centering
  \begin{tikzpicture}
    \pgftext{\pgfimage[width=.75\textwidth]{\KWARCpic{semweb}{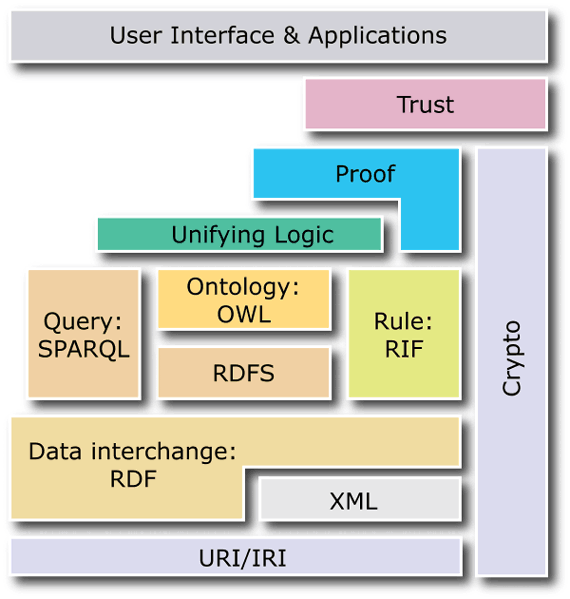}}}
    \node at (3.5,.5) [ellipse,minimum height=1.5cm,fill=red!70!white,draw=white,thick] {\textsf{\textbf{\Large Computation?}}};
  \end{tikzpicture}
  \caption{The Semantic Web Layer Cake (originally by \person{Berners-Lee})}
  \label{fig:layercake}
\end{figure}

As a particularly promising future research task, we have identified the integration of OpenMath-based computations right into queries against RDF-based Linked datasets.  Consider querying a statistical dataset for the region with the highest increase of population density:  Currently, that would require one step of querying (and obtaining population and area values), another step of computation (of the population densities), and a second step of querying (finding the maximum density).  Or reconsider the example of computing a derived value in a dataset from section~\ref{sec:applications}, where an expression is rewritten using the OpenMath definition of a function:  When the arguments of that function are again derived values, we would also have to execute a chain of RDF queries and OpenMath-based term rewritings.  RDF queries are usually made in the SPARQL language~\cite{PruSea08:sparql}.  The SPARQL specification foresees the extension of the basic language by additional entailment regimes~\cite{w3c:sparql-entailment}, which make a query return additional, \emph{entailed} results, beyond the information that is explicitly encoded in the RDF graph being queried.  The possibilities for an OpenMath entailment regime should be investigated.  More pragmatically, and disregarding the consequences for computational complexity, many implementations of SPARQL query processors allow for defining extension functions, and some basic mathematical extension functions have already been implemented for certain query processors; it should be investigated how that can be generalized to arbitrary functions defined by OpenMath CDs.  The goal to be pursued with that is to make \emph{computation} an adequate part of the well-known Semantic Web layer cake (figure~\ref{fig:layercake}).

\paragraph{Acknowledgments:} The original idea for integrating OpenMath CDs into the Web of Data and further steps, such as combining SPARQL and OpenMath, emerged from discussions with \person{Denny Vrandećič} and a joint publication~\cite{VLHBD:SemGovStatData10}.  I would also like to thank \person{David Carlisle}, \person{Jan Willem Knopper} and \person{Michael Kohlhase} for their input.

\printbibliography

\ifdraft
\fi
\end{document}